\documentclass[conference, 10pt]{IEEEtran}

\usepackage{epsfig,amsmath,amssymb,epsf,amsthm,scalefnt,multirow,subfig}
\usepackage{xcolor}
\usepackage{float}
\usepackage{cite}
\usepackage{psfrag}
\usepackage{gensymb}
\usepackage{cleveref}

\newcommand{\argmax}{\mathop{\mathrm{argmax}}}

\newtheorem*{lemma*}{Lemma}

  \def\cC{{\mathcal{C}}}

 \def\cN{{\mathcal{N}}}

\def\argmax{\mathop{\mathrm{argmax}}}

\def\b0{{\pmb{0}}} 

\def\ba{{\mathbf{a}}}   
 \def\bff{{\mathbf{f}}}  
   
 \def\bn{{\mathbf{n}}}  
   
  \def\bw{{\mathbf{w}}}

\def\bA{{\mathbf{A}}}   
 \def\bF{{\mathbf{F}}}  \def\bH{{\mathbf{H}}}
\def\bI{{\mathbf{I}}}

  \def\bW{{\mathbf{W}}}

\begin{document}

\title{Efficient Channel AoD/AoA Estimation Using Widebeams for Millimeter Wave MIMO Systems}

 \author{\IEEEauthorblockN{Hyeongtaek Lee, Sucheol Kim, and Junil Choi}
\IEEEauthorblockA{
	Department of Electrical Engineering\\
	Pohang University of Science and Technology\\
	Pohang, Korea 37673\\
	Email: \{htlee8459, recusmik, junil\}@postech.ac.kr}}

\maketitle

\begin{abstract}
Using millimeter-wave (mmWave) bands is expected to provide high data rates through large licensed and unlicensed spectrum. Due to large path loss and sparse scattering propagation properties, proper beam alignment is important in mmWave systems. Small carrier wavelengths at mmWave bands let wireless communication systems use large antenna arrays to provide sufficient beamforming gain with highly directional beams. Hence, high resolution channel angle-of-departure (AoD) and angle-of-arrival (AoA) estimation is crucial for beam alignment to get the advantages of large beamforming gain. Using large antenna arrays, however, can lead to high system complexity and channel estimation overhead. This paper proposes a channel AoD/AoA estimation technique using widebeams to lower estimation overhead and auxiliary-beam-pair (ABP) to get high resolution channel AoD/AoA estimates considering hybrid transceiver structures. To fully use the hybrid transceiver structures, the linear combination of discrete Fourier transform (DFT) vectors is considered to construct widebeams. Numerical results show that the proposed estimator can get high resolution channel AoD/AoA estimates with lower overhead compared to previous estimators.
\end{abstract}


\section{Introduction}\label{Introduction}
Using millimeter-wave (mmWave) bands is crucial to support high data rates by using abundant licensed and unlicensed spectrum for future wireless communication systems including fifth generation (5G) cellular network \cite{Pi:2011,Heath:2016,Boccardi:2014}. Small carrier wavelengths at mmWave frequencies make it possible to have small array form factors, which support the use of large antenna arrays to generate highly directional beams and provide large beamforming gain. Since the large antenna array makes it difficult to use fully digital multiple-input multiple-output (MIMO) systems, which use one radio frequency (RF) chain for each antenna, many previous works considered hybrid analog and digital precoding and combining techniques as in \cite{Zhang:2005,Roh:2014,Ayach:2014}. 
These hybrid transceiver structures allow to exploit both beamforming gain and spatial multiplexing with reduced hardware cost and complexity.

Unlike classical low-frequency bands, the signal propagation properties at mmWave bands are characterized by large path loss and sparse scattering \cite{Sun:2014}. Hence, to fully use the benefits from beamforming through beam alignment, which is essential for mmWave communications\cite{Choi:2016magazine}, accurate channel knowledge such as channel angle-of-departures (AoDs) and angle-of-arrivals (AoAs) are required at both base station (BS) and user equipment (UE) sides.

As a prior work, grid-of-beams (GoB) based beam training approach was considered in \cite{Singh:2015, Wang:2009, Palacios:2017} to get channel AoD and AoA. In the GoB method, the best combination of transmit and receive beams are chosen through exhaustive or sequential search with respect to the strengths of signals. However, since the estimation performance is defined by the grid resolution, the number of antennas should be increased to generate narrow beams and to reduce the estimation error, which results in high system complexity and estimation overhead.

To tackle the overhead problem, some works in \cite{Hur:2013, Ayach:2014, Alkhateeb:2014, Seo:2016, Wang:2009} considered channel AoD/AoA estimation using hierarchical multi-resolution beam codebooks. The idea is similar to the GoB method but hierarchical codebook approaches first start with widebeams and successively increase the beam resolution by using narrower beams to reduce the estimation overhead. The final beam resolution of the hierarchical codebook approaches, however, is still determined by the number of antennas and the performance of the angle estimation is the same as the GoB method.

In order to obtain high resolution angle estimates, a technique using auxiliary-beam-pairs (ABPs) has been proposed in \cite{Zhu:2017,Zhu:2018}. This technique exploits analog beams to cover the angular range of interest and post-process the power of the signals from these beams to estimate AoD and AoA. Unlike the GoB method, which simply selects the beam with the largest power and sets the boresight of the selected beam as the estimated angle, the ABP technique can estimate the angle even between the boresights of the beams with high resolution. Despite the high resolution, the estimation overhead of the ABP technique is still similar to that of the GoB method.

In this paper, we propose a two-stage channel AoD/AoA estimator for mmWave MIMO systems. At the first stage, a rough possible angular range is obtained through widebeam selection with respect to the strength of signal. Precise angle is estimated using the ABP technique at the second stage based on the knowledge at the first stage. We optimize the structure of widebeams to fully exploit the benefit of the ABP technique. Widebeams with proper beamwidths and boresights are constructed through linear combination of discrete Fourier transform (DFT) vectors, which can, different from the conventional ABP technique that only relies on the analog beamforming, fully exploit the hybrid transceiver structures. The numerical results show that our proposed channel AoD/AoA estimator has high estimation resolution comparable to the conventional ABP technique with much lower overhead.

The rest of this paper is organized as follows. The system model is described in Section \ref{System model}. Then, we propose an angle estimator based on the ABP technique with widebeams in Section \ref{Proposed estimator}. In Section \ref{Numerical result}, we numerically evaluate the proposed estimator and compare with the GoB and ABP techniques. We conclude the paper in Section \ref{conclusion}.

\textbf{Notation:} Lower and upper boldface letters represent column vectors and matrices.  $\bA^{\mathrm{T}}$ and $\bA^\mathrm{H}$ denotes the transpose and conjugate transpose of the matrix $\bA$, and $\lvert\cdot\rvert$ is used to denote the absolute value of the complex number. $\mathbf{0}_m$ is used for the $m\times 1$ all zero vector, and $\bI_m$ denotes the $m \times m$ identity matrix.

\section{System Model}\label{System model}
\begin{figure}
	\centering
	\includegraphics[width=0.95\columnwidth]{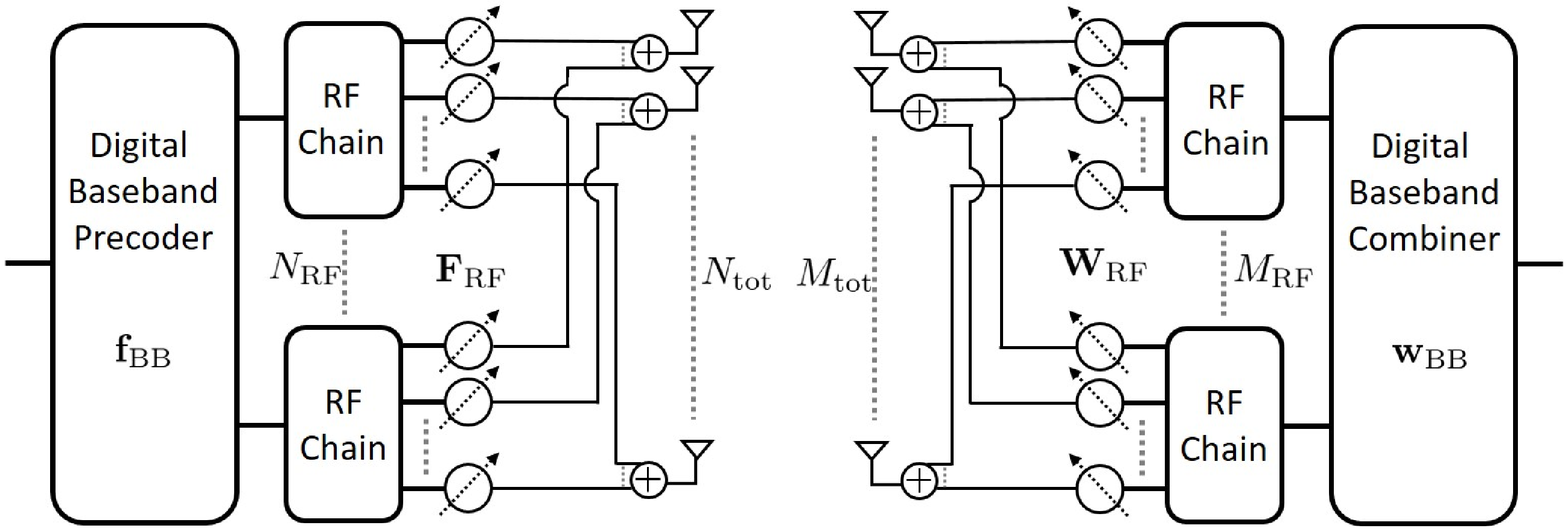}
	\caption{Hybrid transceiver structure of single stream. $N_\mathrm{RF}$ RF chains and $N_\mathrm{tot}$ antennas at the transmitter and $M_\mathrm{RF}$ RF chains and $M_\mathrm{tot}$ antennas at the receiver are assumed.}
	\label{Hybrid system model}
\end{figure}

We consider mmWave MIMO system with a hybrid transceiver structure as shown in Fig. \ref{Hybrid system model}. In the hybrid transceiver structure, in general, it is possible to have spatial multiplexing where the number of data streams can be the minimum of RF chains at the transmitter and receiver. Since we only consider the beam alignment, we assume the baseband precoder and combiner are vectors to perform beamforming.

A transmitter is equipped with $N_\mathrm{tot}$ antennas with $N_\mathrm{RF}$ RF chains for transmitting a data stream to a receiver equipped with $M_\mathrm{tot}$ antennas with $M_\mathrm{RF}$ RF chains. The received symbol $y$ across receive antennas after analog and baseband combining is represented by
\begin{align}\label{input_output1}
y=\sqrt{\rho} \bw_\mathrm{BB}^\mathrm{H} \bW_\mathrm{RF}^\mathrm{H} \bH\bF_\mathrm{RF} \bff_\mathrm{BB} x+\bw_\mathrm{BB}^\mathrm{H} \bW_\mathrm{RF}^\mathrm{H} \bn,
\end{align}
where $x$ is the transmitted symbol assuming $|x|^2=1$, $\bn\sim \cC\cN(\mathbf{0}_{M_{\mathrm{tot}}},\bI_{M_{\mathrm{tot}}})$ is a noise vector and $\rho$ is the signal-to-noise ratio (SNR). $\bF_\mathrm{RF}$ is the $N_\mathrm{tot} \times N_\mathrm{RF}$ analog precoding matrix assuming constant modulus elements and $\bff_\mathrm{BB}$ is the $N_\mathrm{RF} \times 1$ digital baseband precoding vector such that $\lVert \bF_\mathrm{RF}\bff_\mathrm{BB}\rVert_{2}^2=1$. Also, $\bW_\mathrm{RF}$ and $\bw_\mathrm{BB}$ denote $M_\mathrm{tot} \times M_\mathrm{RF}$ and $M_\mathrm{RF} \times 1$ analog and digital baseband combining matrix and vector such that $\lVert\bw_\mathrm{BB}^\mathrm{H}\bW_\mathrm{RF}^\mathrm{H}\rVert_{2}^2=1$, which are defined similar to the precoding matrix and vector.  

We adopt the Rician channel model, and the $M_\mathrm{tot} \times N_\mathrm{tot}$ MIMO channel matrix $\bH$ consists of the line-of-sight (LOS) and non-line-of-sight (NLOS) paths as
\begin{align}\label{rician channel}
\bH=&\sqrt{\frac{K}{1+K}} \underbrace{\alpha_\ell \ba_\mathrm{r}(\phi_\ell) \ba_\mathrm{t}(\theta_\ell)^\mathrm{H}}_{\bH_\mathrm{LOS}} \notag\\
&+\sqrt{\frac{1}{1+K}}\underbrace{\sum_{\ell'\neq\ell}^{L-1} \alpha_{\ell'} \ba_\mathrm{r}(\phi_{\ell'}) \ba_\mathrm{t}(\theta_{\ell'})^\mathrm{H}}_{\bH_\mathrm{NLOS}},
\end{align}
where $K$ is the Rician K-factor, $L$ is the number of total paths, $\alpha_i = g_i \sqrt{N_\mathrm{tot} M_\mathrm{tot}}$ is the complex path gain with $g_i \sim\cC\cN(0,1)$. Also, $\phi_i$ and $\theta_i$ are the AoA and the AoD of the $i$-th path, and $\ba_\mathrm{r}(\cdot)$ and $\ba_\mathrm{t}(\cdot)$ are the receive and transmit antenna array response vectors.

For the algorithm development, we assume the LOS scenario since we are interested in estimating the dominant channel AoD/AoA for the beam alignment. Moreover, due to the strong directivity of mmWave, the LOS assumption is valid in many practical scenarios. In Section \ref{Numerical result}, we consider the Rician channel model for numerical studies. The single path channel matrix is given by 
\begin{align}\label{single path channel}
\bH=\alpha \ba_\mathrm{r}(\phi) \ba_\mathrm{t}(\theta)^\mathrm{H},
\end{align}
where $\phi$ and $\theta$ are the AoA and AoD of the LOS path. We consider a uniform linear array (ULA) at both the transmitter and receiver to exploit key concepts for channel AoD/AoA estimation in \cite{Zhu:2017}. The array response vectors are given by 
\begin{align}\label{ULA}
&\ba_\mathrm{r}(\phi) =\frac{1}{\sqrt{M_\mathrm{tot}}} \left[1, e^{j\frac{2\pi}{\lambda} d_\mathrm{r} \sin(\phi)}, \cdots, e^{j\frac{2\pi}{\lambda}({M_\mathrm{tot}}-1) d_\mathrm{r} \sin(\phi)} \right]^T, \\
&\ba_\mathrm{t}(\theta) = \frac{1}{\sqrt{N_\mathrm{tot}}} \left[1, e^{j\frac{2\pi}{\lambda} d_\mathrm{t} \sin(\theta)}, \cdots, e^{j\frac{2\pi}{\lambda}({N_\mathrm{tot}}-1) d_\mathrm{t} \sin(\theta)} \right]^T,
\end{align}
where $\lambda$ denotes the wavelength, $d_\mathrm{t}$ implies the distance between transmit antenna elements, similarly, $d_\mathrm{r}$ implies the distance between receive antenna elements. Let us denote $\psi=\frac{2\pi}{\lambda}d_\mathrm{r}\sin(\phi)$ and $\mu=\frac{2\pi}{\lambda}d_\mathrm{t}\sin(\theta)$ the transmit and receive spatial frequencies, then the array response vectors in terms of the spatial frequencies are written by
\begin{align}\label{ULA_2}
&\ba_\mathrm{r}(\psi)=\frac{1}{\sqrt{M_\mathrm{tot}}} \left[1, e^{j\psi}, \cdots, e^{j({M_\mathrm{tot}}-1) \psi} \right]^T, \\
&\ba_\mathrm{t}(\mu)=\frac{1}{\sqrt{N_\mathrm{tot}}} \left[1, e^{j\mu}, \cdots, e^{j({N_\mathrm{tot}}-1) \mu} \right]^T.
\end{align}  

\section{Proposed angle estimation based on ABP}\label{Proposed estimator}

In this section, we first describe the two-stage channel AoD/AoA estimator using the ABP technique with widebeams to maximize the estimation accuracy. Then, we explain the method for optimizing the beamwidths of widebeams. The proposed two-stage estimator is not a simple combination of the widebeam and the ABP. To optimize the performance of this combination, the widebeams should be carefully designed considering the ABP structure as detailed in Section III-B.
We only consider the AoD estimation in this section assuming perfect knowledge of the AoA. The AoA estimation can be performed similarly, and the estimate can be fed back to the transmitter using limited feedback as in \cite{Zhu:2017}.

\begin{flushleft}
	A. \textit{Angle estimation using ABP with widebeams}
\end{flushleft}

\begin{figure}
	\centering
	\includegraphics[width=0.9\columnwidth]{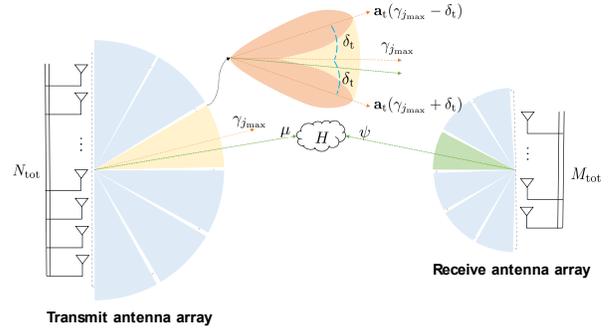}
	\caption{A conceptual example of the proposed AoD estimator.}
	\label{2stage angle estimation}
\end{figure}

Fig. \ref{2stage angle estimation} shows an example of the proposed angle estimator. In the first stage, a rough possible angular range of interest is obtained through an exhaustive search similar to the GoB method but widebeams are used to lower the estimation overhead. Let us denote the precoder of the $j$-th transmit widebeam by $\bff_{\mathrm{WB},j} $ and $J$ by the total number of transmit widebeams. Remark that $\bff_{\mathrm{WB},j}$ is formed by the multiplication of some $\bF_{\mathrm{RF},j}$ and $\bff_{\mathrm{BB},j}$ that will be formally defined in Section III-B. Then the $j$-th received signal is given by
\begin{equation}\label{jth_received}
y_j=\sqrt{\rho}\alpha \ba_{\mathrm{r}}^\mathrm{H}(\psi) \ba_{\mathrm{r}}(\psi) \ba_{\mathrm{t}}^\mathrm{H}(\mu) \bff_{\mathrm{WB},j} x + \ba_{\mathrm{r}}^\mathrm{H}(\psi) \bn.
\end{equation}
In \eqref{jth_received}, we set $\bW_\mathrm{RF}\bw_\mathrm{BB}=\ba_{\mathrm{r}}(\psi)$ based on the perfect AoA knowledge assumption at the transmitter. The corresponding received signal strength can be calculated as $\chi_j=\lvert y_j\rvert^2$.
Since each widebeam has the maximum gain at its boresight and distinct beamwidth, we assume the AoD $\mu$ is assumed to be in the cover range of the $j_\mathrm{max}$-th widebeam where 
\begin{equation}
j_\mathrm{max}=\argmax_{j\in\{1,\dots,J\}}(\chi_j).
\end{equation}

In the second stage, we estimate the AoD $\mu$ precisely based on the ABP technique in \cite{Zhu:2017}. To estimate the angle $\mu$, multiple ABPs can be considered in the coverage of a widebeam depending different parameters, e.g., the beamwidth of widebeams. In this paper, we assume only one ABP is located within the coverage of widebeam for simple illustration.

Let us denote $\gamma_{j_\mathrm{max}}$ by the boresight of the $j_\mathrm{max}$-th widebeam.
Consider an ABP formed by the transmitter as $\ba_{\mathrm{t}}(\gamma_{j_\mathrm{max}}-\delta_{\mathrm{t}})$ and $\ba_{\mathrm{t}}(\gamma_{j_\mathrm{max}}+\delta_{\mathrm{t}})$ where the spatial frequency $\delta_{\mathrm{t}}$ is conditioned by
\begin{equation}\label{delta}
\delta_{\mathrm{t}}=\frac{k\pi}{N_\mathrm{tot}}, \enspace k=1,2,3,\cdots.
\end{equation}
The parameter $k$ will be explained in the next subsection.
Note that the received signals by the ABP is written as
\begin{align}
y^{\Delta}=\sqrt{\rho}\alpha \ba_{\mathrm{r}}^\mathrm{H}(\psi) \ba_{\mathrm{r}}(\psi) \ba_{\mathrm{t}}^\mathrm{H}(\mu) \ba_{\mathrm{t}}(\gamma_{j_\mathrm{max}}-\delta_{\mathrm{t}}) x + \ba_{\mathrm{r}}^\mathrm{H}(\psi) \bn, \\
y^{\Sigma}=\sqrt{\rho}\alpha \ba_{\mathrm{r}}^\mathrm{H}(\psi) \ba_{\mathrm{r}}(\psi) \ba_{\mathrm{t}}^\mathrm{H}(\mu) \ba_{\mathrm{t}}(\gamma_{j_\mathrm{max}}+\delta_{\mathrm{t}}) x + \ba_{\mathrm{r}}^\mathrm{H}(\psi) \bn.
\end{align}
As in \cite{Zhu:2017}, under the large $N_\mathrm{tot} M_\mathrm{tot}$ assumption, the corresponding signal strengths are approximated as
\begin{align} 
\label{chi_delta}
\chi^{\Delta}&=\lvert y^{\Delta} \rvert ^2 \notag \\
&\approx \rho|\alpha|^2  \ba_{\mathrm{t}}^\mathrm{H}(\gamma_{j_\mathrm{max}}-\delta_{\mathrm{t}})\ba_{\mathrm{t}}(\mu)\ba_{\mathrm{t}}^\mathrm{H}(\mu) \ba_{\mathrm{t}}(\gamma_{j_\mathrm{max}}-\delta_{\mathrm{t}}),
\\
\label{chi_sigma}
\chi^{\Sigma}&=\lvert y^{\Sigma} \rvert ^2 \notag \\ 
&\approx \rho|\alpha|^2  \ba_{\mathrm{t}}^\mathrm{H}(\gamma_{j_\mathrm{max}}+\delta_{\mathrm{t}})\ba_{\mathrm{t}}(\mu)\ba_{\mathrm{t}}^\mathrm{H}(\mu) \ba_{\mathrm{t}}(\gamma_{j_\mathrm{max}}+\delta_{\mathrm{t}}), 
\end{align}
The signal strengths in \eqref{chi_delta} and \eqref{chi_sigma} can be represented as
\begin{align}
\chi^{\Delta}=\rho|\alpha|^2 \frac{\cos^2\left(\frac{N_{\mathrm{tot}}\left(\mu-\gamma_{j_\mathrm{max}}\right)}{2}\right)}{\sin^2\left( \frac{\mu-\gamma_{j_\mathrm{max}}+\delta_{\mathrm{t}}}{2}\right)},
\end{align}
\begin{align}
\chi^{\Sigma}=\rho|\alpha|^2 \frac{\cos^2\left(\frac{N_{\mathrm{tot}}(\mu-\gamma_{j_\mathrm{max}})}{2}\right)}{\sin^2\left( \frac{\mu-\gamma_{j_\mathrm{max}}-\delta_{\mathrm{t}}}{2}\right)},
\end{align}
and the ratio metric is defined by
\begin{equation}
\zeta^{\mathrm{AoD}}=\frac{\chi^{\Delta}-\chi^{\Sigma}}{\chi^{\Delta}+\chi^{\Sigma}}=-\frac{\sin(\mu-\gamma_{j_\mathrm{max}})\sin(\delta_{\mathrm{t}})}{1-\cos(\mu-\gamma_{j_\mathrm{max}})\cos(\delta_{\mathrm{t}})}.
\end{equation}
By the monotonicity of the ratio metric shown in \cite{Zhu:2017}, the estimated value of $\mu$ is derived by
\begin{align}\label{ratio metric}
&\hat{\mu}=\gamma_{j_\mathrm{max}}- \notag \\ 
&\arcsin\left(\frac{\zeta^{\mathrm{AoD}}\sin(\delta_{\mathrm{t}})-\zeta^{\mathrm{AoD}}\sqrt{1-(\zeta^{\mathrm{AoD}})^2}\sin(\delta_{\mathrm{t}})\cos(\delta_{\mathrm{t}})}{\sin^2(\delta_{\mathrm{t}})+(\zeta^{\mathrm{AoD}})^2\cos^2(\delta_{\mathrm{t}})}\right).
\end{align}

\begin{flushleft}
	B. \textit{Adequate widebeam generation}
\end{flushleft}

Unlike the GoB and the hierarchical codebook based methods, which uniformly sectorize the considering angular range, we optimize the beamwidth of widebeams to maximize the estimation performance of proposed estimator. As assumed in Section III-A, the two analog beams consisting the ABP should be $2\delta_{\mathrm{t}}$ apart from each other in the sense of their boresight. Also, the spatial frequency $\delta_{\mathrm{t}}$ has the specific condition \eqref{delta}. Since the selected widebeam that has the maximum signal strength is changed by the channel AoD, the beamwidth of widebeams should be optimized to set the two analog beams on the edges of the selected widebeam. Therefore, we call adequate widebeam as the widebeam where its beamwidth is $2\delta_{\mathrm{t}}$ for some positive integer $k$ in \eqref{delta}. The appropriate $k$ is chosen according to the target beamwidths of the widebeams.

To generate adequate widebeams, we need to easily control the beamwidths of widebeams. We adapt the variation of the method in \cite{Noh:2017}, which uses the linear combination of the DFT vectors to design a multi-resolution codebook. The resulting widebeam precoder is written as
\begin{align}
\bff_{\mathrm{WB},j}&=\bF_{\mathrm{RF},j}\bff_{\mathrm{BB},j} \\
\bF_{\mathrm{RF},j}&=[\ba_{\mathrm{t}}(\gamma_j), \ba_{\mathrm{t}}(\gamma_j+\xi_1),
\ba_{\mathrm{t}}(\gamma_j-\xi_1),\notag \\ &\ba_{\mathrm{t}}(\gamma_j+\xi_2),\ba_{\mathrm{t}}(\gamma_j-\xi_2)\cdots],\\
\bff_{\mathrm{BB},j}&=\left[ c_0,c_1,c_1^*,c_2,c_2^*,\cdots \right]^T
\end{align}
where $\gamma_j$ is the boresight of the precoder $\bff_{\mathrm{WB},j}$ and $\xi_i\in[0, 2\pi)$ and $c_i\in\mathbb{C}$ with $i\in\left\{ 1,\cdots,\lfloor \frac{N_\mathrm{RF}-1}{2} \rfloor \right\}$ are free variables to be optimized to get a proper beamwidth. The columns of the  $\bF_{\mathrm{RF},j}$ are the array response vectors where the spatial frequencies are symmetrically chosen to have near symmetricity of the main lobe. To have the maximum gain at the boresight, the components of the $\bff_{\mathrm{BB},j}$ are conditioned by
\begin{align} 
|c_0| \ge |c_1| \ge |c_2| \cdots \ge 0,
\end{align}
where this condition also contributes to the symmetricity of the main lobe. Unlike the conventional ABP technique in \cite{Zhu:2017} that only exploits the analog precoding, the linear combination of the DFT vectors to generate widebeams let us fully use the hybrid transceiver structures.


\begin{figure}
	\centering
	\includegraphics[width=0.9\columnwidth]{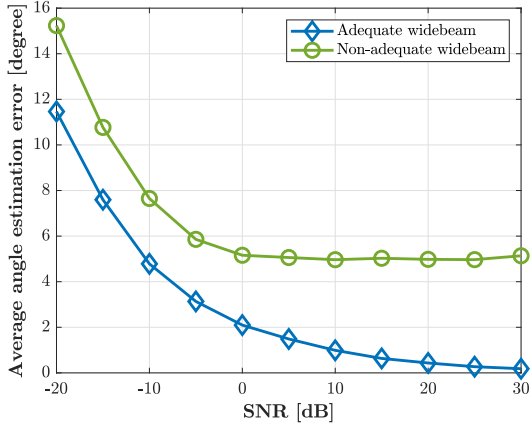}
	\caption{Average angle estimation error difference between adequate and non-adequate widebeams with $N_\mathrm{tot}=16$ assuming the single path channel.}
	\label{ade vs non-ade}
\end{figure}

\section{Numerical Results}\label{Numerical result}

Here, we perform simulations based on the Monte-Carlo approach to evaluate the performance of the proposed two-stage angle estimator. We use average angle (especially the AoD) estimation error as the performance metric, which is $\epsilon_\theta=\mathbb{E}\left[\lvert\theta-\hat{\theta}\rvert\right]$ where $\theta$ and $\hat{\theta}$ implies the true AoD and its estimate. We set $M_\mathrm{tot} = 8$ and assume perfect knowledge of the AoA $\phi$ for all cases. The true $\theta$ and $\phi$ are assumed to be uniformly distributed in [-50\degree, 50\degree] and [-90\degree, 90\degree] each. 

\begin{figure}
	\centering
	\includegraphics[width=0.9\columnwidth]{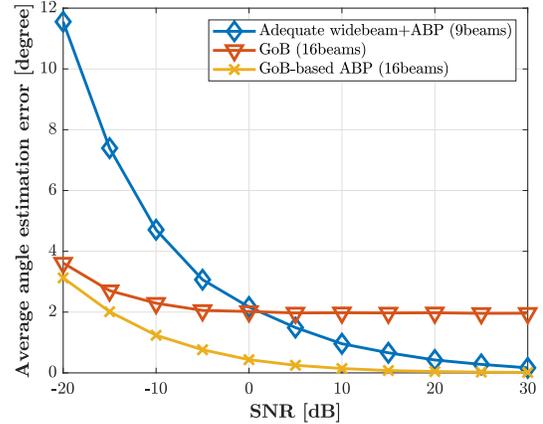}
	\caption{Comparison between proposed estimator and the GoB, GoB-based ABP methods with $N_\mathrm{tot}=16$ assuming the single path channel.}
	\label{N=16}
\end{figure}

\begin{figure}
	\centering
	\includegraphics[width=0.9\columnwidth]{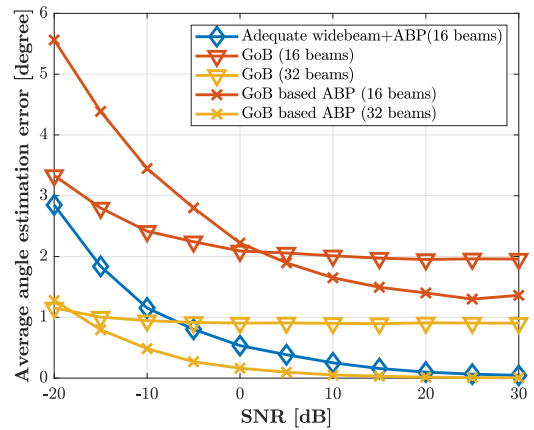}
	\caption{Comparison between proposed estimator and the GoB, GoB-based ABP with $N_\mathrm{tot}=32$ assuming the single path channel.}
	\label{N=32}
\end{figure}

In Figs. \ref{ade vs non-ade}, \ref{N=16} and \ref{N=32}, we assume the singe path channel represented in \eqref{single path channel}. Fig. \ref{ade vs non-ade} shows the performance gap between adequate widebeams and non-adequate widebeams. Non-adequate widebeams refer to the widebeams that the beamwidth is still $2\delta_{\mathrm{t}}$ but does not satisfy \eqref{delta}. The figure shows noticeable performance degradation with saturation in high SNR regime when using the non-adequate widebeams. Then, in the following figures, we compare our proposed angle estimator with the GoB and GoB-based ABP methods, which only use analog beams. The GoB method is nothing but uniformly sectorizing the AoD region. The GoB-based ABP method executes angle estimation using the ABP technique with analog beams covering the sectorized AoD region like the GoB method.

In Fig. \ref{N=16}, the average angle estimation error of the proposed estimator, the GoB and the GoB-based ABP methods are evaluated. We set $N_\mathrm{tot}=16$ in this case. Our proposed estimator uses 7 beams to cover the AoD region and 2 beams to use the ABP technique, which results in the total number of beams used for estimation is 9. The GoB and GoB-based ABP methods both use 16 analog beams each to cover the same AoD region. The proposed estimator uses less number of beams to lower the estimation overhead; however, it outperforms the GoB method from moderate SNR values and achieves almost the same estimation accuracy with the GoB-based ABP method in the high SNR regime.

In Fig. \ref{N=32}, we set $N_\mathrm{tot}=32$ and compare different numbers of analog beams for the GoB and GoB-based ABP methods. Our proposed estimator uses 16 beams (14 beams to cover the AoD region and 2 beams for the ABP) while the other two methods use 16 or 32 beams. When using the same number of beams, the proposed estimator outperforms the other two competitors, and the gap is quite significant even in the high SNR regime. Comparing the proposed estimator with the other two methods using 32 beams, the proposed estimator has lower overhead but shows lower average estimation error than the GoB methods and almost the same performance with the GoB-based ABP method as SNR increases. Figs. \ref{N=16} and \ref{N=32} show that our proposed estimator is more efficient than the others.

\begin{figure}
	\centering
	\includegraphics[width=0.9\columnwidth]{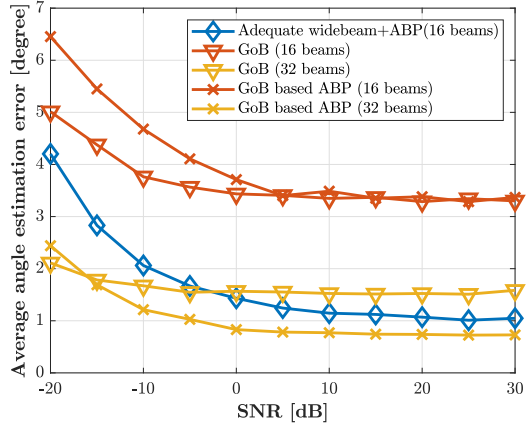}
	\caption{Comparison between proposed estimator and the GoB, GoB-based ABP with $N_\mathrm{tot}=32$ assuming the Rician channel.}
	\label{rician_N=32}
\end{figure}

In Fig. \ref{rician_N=32}, we consider the Rician channel in \eqref{rician channel} assuming $K=13.5 \text{ }\mathrm{dB}$ and $L=4$. The object is to estimate the AoD of the dominant path. 
Similar to the single path scenario, $\phi_i$ and $\theta_i$ are assumed to be uniformly distributed in the same angular regions where all other parameters are the same as in Fig. \ref{N=32}. 
The overall trend is similar to that of Fig. \ref{N=32}; however, the NLOS paths are treated as noise resulting in saturation in the high SNR regime. Fig. \ref{rician_N=32} still shows that proposed two-stage estimator is more efficient even for the multipath scenario.
\section{Conclusion}\label{conclusion}
In this paper, we proposed the two-stage channel AoD/AoA estimator using the ABP with widebeams. At the first stage, which carries out widebeam selection to get a rough possible angular range of interest, we decreased estimation overhead by using adequate widebeams. More precise angle is estimated at the second stage using the ABP technique. We optimized the structure of widebeams to fully exploit the advantages of the ABP technique and the hybrid transceiver structures. The numerical results showed that our proposed estimator is more efficient than the GoB and GoB-based ABP methods for both single and multipath scenarios.

\section{Acknowledgment}

This work was partly supported by Institute for Information \& communications Technology Promotion (IITP) grant funded by the Korea government (MSIT) (No. 2016-0-00123, Development of Integer-Forcing MIMO Transceivers for 5G \& Beyond Mobile Communication Systems) and the National Research Foundation (NRF) grant funded by the MSIT of the Korea government (No. 2019R1C1C1003638).

\bibliographystyle{IEEEtran}
\bibliography{refs_all}

\end{document}